%% file: main.tex
\documentstyle[preprint]{ptptex}

\title{The Intersection Angles between $N$-Dimensional Stable and
  Unstable Manifolds in $2N$-Dimensional Symplectic Mappings}
\author{Yoshihiro HIRATA,\thanks{E-mail address:
    yhirata@allegro.phys.nagoya-u.ac.jp} Kazuhiro NOZAKI and Tetsuro
  KONISHI}
\inst{Department of Physics, Nagoya University, Nagoya, 464-8602,
  Japan}
\abst{\input{abstract.tex}}
\preprintnumber{DPNU--99--20\\June 1999}
\notypesetlogo

\input{Greek.tex}

\input{NdimMap.tex}

\input{math.tex}

\input{general.tex}

\begin{document}

\maketitle

\input{section}

\input{bib}
\end{document}

%% file: abstract.tex
We asymptotically compute the intersection angles between
$N$-dimensional stable and unstable manifolds in $2N$-dimensional
symplectic mappings.
There exist particular $1$-dimensional stable and unstable
sub-manifolds which experience exponentially small splitting of
separatrix in our models.
We show that the angle between the sub-manifolds is exponentially
small with respect to the perturbation parameter $\e$, and the other
angles are $\O{\e^2}$.

%% file: Greek.tex
\def\a{\alpha}

\def\e{\epsilon}

\def\aj{\alpha_j}

\def\bj{\beta_j}



%% file: NdimMap.tex
\def\vecq{\vec{q}}
\def\vecp{\vec{p}}
\def\vecg{\vec{\gamma}}

\def\pj{p_j}
\def\qj{q_j}

\def\pjp{p_j^{\sss \prime}}
\def\qjp{q_j^{\sss \prime}}

\def\Vzero{V_0}
\def\Vone{V_1}

\def\aj{a_j}
\def\bj{b_j}

\def\qjpu{q_{j+1}}

\def\fzero{f_0}
\def\fonej{f_{j,1}}

\def\tj{t_j}

%% file: math.tex
\def\R{{\bf R}}

\def\det#1{{\rm det}#1}

\def\del{\partial}

\def\sech#1{{\rm sech}(#1)}
\def\sechd#1{{\rm sech}^2(#1)}
\def\secht#1{{\rm sech}^3(#1)}

\def\sinh#1{{\rm sinh}(#1)}
\def\sinhd#1{{\rm sinh}^2(#1)}

\def\O#1{O(#1)}

\def\exp#1{e^{#1}}
\def\Exp#1{{\rm exp}\left(#1\right)}

\def\atan#1{{\rm tan}^{-1}#1}

\def\acos#1{{\rm cos}^{-1}#1}

\makeatletter
\def\@begintheorem#1#2{\rm \trivlist
     \item[\hskip \labelsep{\bf #1\ #2}] \item[]}
\def\@opargbegintheorem#1#2#3{\rm \trivlist
     \item[\hskip \labelsep{\bf #1\ #2\ (#3)}] \item[]}
\makeatother

%% file: general.tex
\def\sss{\scriptscriptstyle}

\def\ds{\displaystyle}



%% file: section.tex
The existence of a transversal intersection between stable and
unstable manifolds leads to non-integrability of a Hamiltonian
system.
Besides, in 2-dimensional symplectic mappings (area-preserving
mappings), which correspond to Poincar\'e mappings of Hamiltonian
systems with 2 degrees of freedom, the area of the lobes enclosed by
stable and unstable manifolds stands for flux from inside
(resp. outside) of separatrix to outside (resp. inside).
Because the intersection angle at the principal intersecting point is
related with the area of the lobes, the intersection angle is also
related with flux.
Therefore the intersection angle is essentially important to
understand flux near separatrix and, probably, global transport of
phase space.

%
For some 2-dimensional symplectic maps, e.g. the standard map, H\'enon
map, and the double-well map, one can analytically construct
asymptotic expansions of stable and unstable
manifolds.\cite{HM93,TTJ94,TTJ98,NH96}
Hence one can also compute the asymptotic form of the intersection
angles between stable and unstable manifolds in such maps.
We have been extending the method to 4- or more-dimensional symplectic 
maps with two or more hyperbolic modes, to which Melnikov's
method\cite{GH} cannot be applied.

Such an approach was first presented by Gelfreich and
Sharomov\cite{GS95} and we generalized their model in
Ref.~\citen{HK}.
In addition, we constructed the asymptotic expansions of particular
$1$-dimensional stable and unstable sub-manifolds which experience
{\it exponentially small splitting of separatrix}, in a 4-dimensional
double-well symplectic mapping.
Furthermore, the neighborhood of the sub-manifolds was asymptotically
studied and exponentially small oscillating terms were successfully
obtained.\cite{HNK99}
In this paper we treat $2N$-dimensional double-well symplectic
mappings and standard mappings with nearest-neighbor two-body
couplings and asymptotically compute the intersection angles between
$N$-dimensional stable and unstable manifolds.

The canonical coordinates we use are $\vecq = (q_1, \cdots, q_N) \in
\R^N$, $\vecp = (p_1, \cdots, p_N) \in \R^N$ and $\vecg = (\vec{q},
\vec{p})$.
We start with $2N$-dimensional symplectic mappings $\vecg \mapsto
\vecg^{\sss \prime}$,
\begin{eqnarray}
  \label{eq:map0}
  \left\{
    \begin{array}{rcl}
      \pjp & = & \ds \pj - \e \frac{\del \Vzero (\qj )}{\del \qj} - \e^3
      \frac{\del \Vone (\vecq )}{\del \qj}, \\
      \qjp & = & \qj + \e \pjp,
    \end{array}
  \right.
\end{eqnarray}
where $j = 1, 2, \cdots, N$,\footnote{We shall use the subscript $j$
  as meaning $j = 1, 2, \cdots, N$ without further notice.} and $|\e| 
\ll 1$ is a perturbation parameter.
We take account of two different mappings:
One is the double-well mapping (DW), the other is the standard mapping 
(SM).
The potential function $\Vzero (\qj )$ is given as 
$\ds \frac{1}{4}q^4 - \frac{1}{2}q^2$ (for DW), $\cos{q}$ (for SM),
respectively.
For both of the potential functions, the origin $(0, \cdots, 0)$ is a
$2N$-dimensional hyperbolic fixed point (a direct product of $N$
$2$-dimensional hyperbolic fixed points), and there exist
$N$-dimensional stable manifolds and $N$-dimensional unstable
manifolds near the origin.
The coupling potential $\Vone (\vecq )$ is an even polynomial of the
distance between nearest-neighbors of order 4, i.e.,
$\ds \Vone (\vecq ) = \sum_{j=1}^{N}\frac{\aj}{2} ( \qj - \qjpu )^2 +
\sum_{j=1}^{N}\frac{\bj}{4} ( \qj - \qjpu )^4$, where $\aj$'s and
$\bj$'s are real constants and $q_{{\sss N}+1} = q_1$.
In this paper we restrict $a_1 = a_2 = \cdots = a_N = a$ and $b_1 =
b_2 = \cdots = b_N = b$ to simplify the following computation.

Setting $q_j (t) = q_j$ and $q_j(t+\e ) = \qjp$, one can rewrite the
map (\ref{eq:map0}) into the second order difference equation
\begin{eqnarray}
  \label{eq:DifferenceEq}
  \Delta^2_\e \qj (t) & = & \fzero (\qj ) + \e^2 \fonej (\vecq )
\end{eqnarray}
where $\Delta^2_\e q(t) = \{ q(t+\e ) - 2q(t) + q(t-\e ) \}/\e^2$,
$\ds \fzero (\qj) = - \frac{\del \Vzero (\qj )}{\del{\qj}}$ and $\ds
\fonej (\vecq ) = - \frac{\del \Vone (\vecq )}{\del \qj}$.
By regarding the independent variable $t$ as continuous one, the
difference equation (\ref{eq:DifferenceEq}) expresses time evolution
of not trajectories but invariant manifolds, because the set of
trajectories between $t$ and $t + \e$ can be took into consideration
in (\ref{eq:DifferenceEq}).
Expanding $\ds \qj (t \pm \e ) = \sum_{l=0}^{\infty} {\frac{(\pm \e
    )^l}{l!} \frac{d^l\qj}{dt^l}}$, one obtains
\begin{eqnarray}
  \label{eq:ODE}
  \frac{d^2\qj}{dt^2} & = & \fzero (\qj ) + \e^2 \fonej (\vecq ) - 2
  \sum_{l=2}^{\infty} { \frac{\e^{2l-2}}{(2l)!}
    \frac{d^{2l}\qj}{dt^{2l}}},
\end{eqnarray}
which we call the outer equation.

The stable and unstable solutions $q_j^{\pm}(t;\e )$ are asymptotic to
the origin as $t \to \pm\infty$, respectively.
Hence the boundary condition with which we solve the difference
equation (\ref{eq:DifferenceEq}) is given as $q_j^\pm (t;\e ) \to +0$
(for DW), $\mp 0$ (for SM) as $t$ tends to $\pm\infty$.

We construct the stable and unstable solution to the outer equation
(\ref{eq:ODE}) by expanding $\qj^\pm (t;\e )$ into the power series of
$\e^2$, i.e., $\qj^\pm (t,\e ) = q_{j,0}^\pm (t) + \e^2 q_{j,1}^\pm
(t) + \cdots$.
In the same way, the stable and unstable manifolds $\vec{\gamma}^\pm$
is expanded as
\begin{eqnarray}
  \label{eq:manifoldexpansion}
  \vec{\gamma}^\pm & = & \vec{\gamma}_0^\pm + \e^2 \vec{\gamma}_1^\pm
  + \cdots.
\end{eqnarray}

By substituting the expansion into the outer equation (\ref{eq:ODE}),
one gets the unperturbed equation as
\begin{eqnarray}
  \label{eq:ODEup}
  \frac{d^2 q_{j,0}^\pm}{dt^2} & = & \fzero (q_{j,0}^\pm)
\end{eqnarray}
and the perturbed equation (the linearized equation) as
\begin{eqnarray}
  \label{eq:ODEp}
  \frac{d^2 q_{j,l}^\pm}{dt^2} & = & \fzero^{\sss \prime} (q_{j,0}^\pm 
  ) q_{j,l}^\pm + \tilde{f}_{j,l}(\vec{q}^\pm )
\end{eqnarray}
for $l = 1, 2, \cdots$, where the inhomogeneous term $\tilde{f}_{j,l}$ 
is a polynomial of $q_{j,0}^\pm (t)$'s and their derivatives,
e.g., $\ds \tilde{f}_{j,1} = \fonej (\vec{q}_0^\pm ) - \frac{1}{12}
\frac{d^4 q_{j,0}^\pm}{dt^4}$.
In the same way the boundary condition is also expanded as
$q_{j,l}^\pm \to +0$ (for DW), $\mp 0$ (for SM) as $t \to
\pm\infty, \, l = 0,1,\cdots$.
Note that the ordinary differential equations for $q_{j,l}, \, l =
0,1,\cdots$ form a Hamiltonian system with $N$ degrees of freedom in
order-by-order.

The unperturbed solution $q_{j,0}^\pm (t)$ is obtained as $q_{j,0}^\pm
= s(t+\tj )$ where $s(t) = \sech{t}$ (for DW), $4 \atan{(\exp{t})}$
(for SM),
and $t_j$'s are constants of integration.
Let us put $s(t+\tj ) = s_j(t)$ to simplify notations.
The unperturbed solution $s_j(t)$ does not depend on the sign $\pm$,
i.e. the stable and unstable solutions behave completely same for
$-\infty < t < \infty$ and intersect tangently.
That is a manifestation of integrability of the unperturbed equation
(\ref{eq:ODEup}).

By taking $N$ of $t_j$'s and $t$ as independent parameters,
$N$-dimensional stable and unstable manifolds can be constructed.
Because the map (\ref{eq:map0}) is autonomous, it is translationally
invariant with respect to $t$.
Hence one of $t_j$'s is not essential and can be eliminated.
This point will be argued later.

To construct the solutions to the linearized equation (\ref{eq:ODEp}), 
we first consider the homogeneous equations, $\ds \frac{d^2
  q_{j,l}^\pm}{dt^2} = \fzero^{\sss \prime} (q_{j,0}^\pm ) q_{j,l}$,
which are decoupled into $N$ independent equations.
The fundamental system of solutions to the homogeneous equation is
$\dot{s}_j(t) = \dot{s}(t+\tj )$ and $g_j(t) = g(t+\tj )$
where $\ds g(t) = \dot{s}\int{\frac{dt}{\dot{s}^2}}\,$.
The stable and unstable solutions to the equation (\ref{eq:ODEp}) can 
be written as
\begin{eqnarray}
  \label{eq:Psol}
  q_{j,1}^\pm (t,\vec{t}_j) & = & g_j(t) \int_{\pm\infty}^{t}
  {\dot{s}_j(t) \tilde{f}_{j,1}(t) dt} - \dot{s}_j(t) \int_{0}^{t}
  {g_j(t) \tilde{f}_{j,1}(t) dt},
\end{eqnarray}
respectively.

When $\vec{t}_j = 0$, the r.h.s. of the expression (\ref{eq:Psol}) can 
be easily integrated.
The result does not depend on $j$ and the sign $\pm$, hence putting
$q_1(t) = q_{j,1}^\pm (t,0)$, one obtains $\ds q_1 (t) = \frac{1}{3}
\secht{t} - \frac{7}{24} \sech{t} + \frac{t}{24} \sinh{t} \sechd{t}$
(for DW), $\ds \frac{1}{4} \sinh{t} \sechd{t} - \frac{t}{12} \sech{t}$ 
(for SM).
The solution $q_j^\pm (t,0) = s(t) + \e^2 q_1 (t) + \O{\e^4}$ stands
for the $1$-dimensional stable and unstable sub-manifolds, which we
call {\it separatrix}.
Note that the sub-manifolds experience exponentially small splitting
and split into two different manifolds in the order of $\ds \Exp
{ -\frac{\pi^2}{\e}}$.\cite{HK}

To simplify the further computation we transform $t_j$'s into new
parameters $\a_j$'s as $(t_1,\cdots,t_N) = (\a_1,\cdots,\a_N) M$,
where
\begin{eqnarray}
  \label{eq:LinTrans}
  M & = & \left(
    \begin{array}{cccc}
      -1 & 1 & & \\
      & -1 & \ddots & \\
      & & \ddots & 1\\
      & & & -1
    \end{array}
  \right).
\end{eqnarray}
Because $\det{M} = (-1)^N \ne 0$, the transformation is
non-degenerate.
We put $\alpha_N = 0$ without loss of generality.
The $N$-dimensional stable and unstable manifolds are spanned with
these $N-1$ parameters, $\alpha_1, \cdots, \alpha_{N-1}$ and $t$.

The tangent spaces of the sub-manifolds are necessary to estimate the
intersection angles on the sub-manifolds.
We write the stable and unstable manifolds as $\vec{\gamma}^{\pm}
(t,\vec{\alpha}_j) = ( \vec{q}^{\pm} (t,\vec{\alpha}_j), \vec{p}^{\pm}
(t,\vec{\alpha}_j) )$, respectively.
Let us suppose that the stable and unstable manifolds intersect at a
point $z_0 = \vec{\gamma}^{\pm}(t_0,0)$ where $t_0 \in \R$.
We write tangent vectors of the manifolds at the point as
$\vec{X}^{\pm}_j$, where $\ds \vec{X}^{\pm}_k = \frac{\del
  \vec{\gamma}^\pm}{\del \alpha_k} (t_0,0),\, k = 1,\cdots,N-1$ and
$\ds \vec{X}^{\pm}_N = \frac{\del \vec{\gamma}^\pm}{\del t} (t_0,0)$.
We write $\vec{X}^\pm_k = ( \vec{x}^\pm_k (t_0), \vec{v}^\pm_k (t_0)
)$, where $\ds \vec{v}^\pm_k (t) = \frac{d\vec{x}^\pm_k}{dt}$, and
$\vec{x}^\pm_k (t)$ can be expanded with respect to $\e$ as
$\vec{x}^\pm_k (t) = \vec{x}^\pm_{0,k} (t) + \e^2 \vec{x}^\pm_{1,k}
(t) + \O{\e^4}$.
Owing to $\vec{\gamma}_0 = (s_1(t), \cdots, s_N(t), \dot{s}_1(t),
\cdots, \dot{s}_n(t))$ (see the expression
(\ref{eq:manifoldexpansion}) and the discussion below it), one obtains
\begin{eqnarray}
  \label{eq:q0k}
  \vec{x}_{0,k} & = & \left.\frac{\del}{\del \alpha_k} (s_1(t),
  \cdots, s_N(t) )\right|_{\vec{\alpha} = 0} \nonumber \\
  & = & (0, \cdots, 0, \stackrel{k}{\breve{-\dot{s}}},
  \stackrel{k+1}{\breve{\dot{s}}}, 0, \cdots, 0), \mbox{ for $k = 1,
  \cdots, N-1$} \nonumber \\
  \vec{x}_{0,N} & = & (\dot{s}, \cdots, \dot{s}).
\end{eqnarray}
Note that $\vec{X}^{\pm}_k \perp \vec{X}^{\pm}_N,\, \vec{X}^{\pm}_k
\perp \vec{X}^{\pm}_{k+m},\, k = 1,\cdots,N-1, \, m \ge 2$, and
$\vec{X}^{\pm}_k \not\perp \vec{X}^{\pm}_{k+1},\, k \ne N-1$ up to
$\O{\e^2}$.

The set of tangent vectors $\ds \{ \vec{X}_j^\pm \}$ is a basis of the
tangent space of the stable and unstable manifolds at $z_0$,
respectively.
We denote the inner product between $\vec{X}_m^+$ and $\vec{X}_n^-$ as
$\langle \vec{X}_m^+ | \vec{X}_n^- \rangle$ where $1 \le m,n \le N$.
The intersection angles between the stable and unstable manifolds at
$z_0$ are defined as
\begin{eqnarray}
  \label{eq:ThetaMN}
  \theta_{m,n} (t_0) & = & \acos{\frac{\langle \vec{X}_m^+ |
      \vec{X}_n^- \rangle}{|\vec{X}_m^+| |\vec{X}_n^-|}}.
\end{eqnarray}
Especially, $\theta_{N,N}$ stands for {\it exponentially small
  splitting of separatrix},\cite{HM93,NH96,LST89} i.e., there exist
  $C > 0$ such that
\begin{eqnarray}
  \label{eq:thetaNNgive}
  | \theta_{N,N} (t_0) | & \le &
      \ds \frac{C}{\e^\sigma} \Exp{-\frac{\pi^2}{\e}},
\end{eqnarray}
where $\sigma = 5$ (for DW), 3 (for SM),
at the principal intersecting point ($t_0 = 0$).

To compute the intersection angles $\theta_{m,n}$, one has to estimate 
$\vec{X}^\pm_j$ including the terms of, at least, $\O{\e^2}$.
Hence we shall analyze the neighborhood of the separatrix as a
perturbation problem with respect to $\a_j$'s.
We expand $q_{j,1}^\pm (t,\vec{\a})$ (see the expression
(\ref{eq:Psol})) with respect to $\a_j$'s as
\begin{eqnarray}
  \label{eq:ExpAlpha}
  q_{j,1}^\pm (t,\vec{\a}) & = & q_1(t) + \sum_{k=0}^{N-1} {\a_k
    \tilde{q}_{j,k}^\pm (t)} + \O{|\vec{\a}|^2},
\end{eqnarray}
where $\ds \tilde{q}_{j,k}^\pm (t) = \left. \frac{\del q_{j,1}^\pm
    (t,\vec{\a})}{\a_k} \right|_{\vec{\a} = 0}$.
From the expression (\ref{eq:Psol}), straightforward computation gives
\begin{eqnarray}
  \label{eq:TildeQjk}
  \tilde{q}_{j,k}^\pm (t,\vec{\alpha}) & = & t_{j,k} F_1^\pm (t) +
  ( t_{j+1,k} + t_{j-1,k} ) F_2^\pm (t),
\end{eqnarray}
where $\ds t_{j,k} = \frac{\del t_j}{\del \a_k}$.
The functions $F_1^\pm(t)$ and $F_2^\pm(t)$ is explicitly written by
using the functions $\dot{s}(t)$ and $g(t)$ as
\begin{eqnarray}
  F_1^\pm (t) & = & \dot{q}_1 (t) \pm a \beta g(t) - a \delta
  \dot{s}(t)\, \sinhd{t}, \nonumber \\
  \label{eq:Fone}
  F_2^\pm (t) & = & \mp \frac{1}{2} a \beta g(t),
\end{eqnarray}
where $\ds \beta = \int_{-\infty}^{\infty} \dot{s}(t)^2 dt =
\frac{2}{3}\, ({\rm DW}),\, 8\, ({\rm SM})$, and $\ds \delta =
\frac{1}{3}\, ({\rm DW}),\, 1\, ({\rm SM})$.
The detail of this computation will be reported elsewhere.\cite{HNK}
Note that the expression (\ref{eq:Fone}) contains the parameter $a$,
which determines the strength of coupling.

%
We are now ready to give the explicit representation of
$\vec{X}^\pm_j$ including the terms of $\O{\e^2}$.
Because $\vec{\gamma}_1 = ( \tilde{q}_{1,1}^\pm, \cdots,
\tilde{q}_{N,1}^\pm, \tilde{p}_{1,1}^\pm, \cdots, \tilde{p}_{N,1}^\pm
)$ where $\ds \tilde{p}_{j,1}^\pm = \frac{d \tilde{q}_{j,1}^\pm}{dt}$,
by using the expression (\ref{eq:TildeQjk}), one can explicitly write
$\vec{x}^\pm_{1,k}$ as
\begin{eqnarray}
  \label{eq:TanVec}
  \vec{x}_{1,k}^\pm  & = & ( \tilde{q}_{1,1}^\pm, \cdots,
  \tilde{q}_{N,1}^\pm ) \nonumber \\
  & = & (0, \cdots, 0, \stackrel{k-1}{\breve{{-F_2^\pm} }},
  \stackrel{k}{\breve{F_2^\pm - F_1^\pm }},
  \stackrel{k+1}{\breve{F_1^\pm  - F_2^\pm }},
  \stackrel{k+2}{\breve{F_2^\pm }}, 0, \cdots, 0),
\end{eqnarray}
for $k = 1, \cdots, N-1$.\footnote{The index of the elements is taken
  periodically and overlapped elements (if exist) are summed. For
  example, when $N = 2$, $\vec{x}_{1,1}^\pm = ( -F_1^\pm + 2F_2^\pm,
  F_1^\pm - 2F_2^\pm )$ by definition and the expression
  (\ref{eq:TildeQjk}).
  The first element is regarded as the sum of the elements of index
  $k$ and $k+2$ ($k \equiv k+2\, mod\, 2$), and in the same way,
  the second element is the sum of index $k-1$ and $k+1$.}
By substituting the expressions (\ref{eq:q0k}) and (\ref{eq:TanVec})
into the equation (\ref{eq:ThetaMN}) and using the expression
(\ref{eq:Fone}), one obtains the intersection angles between the
stable and unstable manifolds at the point $z_0$ as
\begin{eqnarray}
  \label{eq:angle}
  \theta_{k,k} (t_0) & = & \frac{\e^2 |\beta|}{\dot{s}(t_0)^2 +
    \ddot{s}(t_0)^2}\, [\, 9 + ( g(t_0)^2 + \dot{g}(t_0)^2 )
  ( \dot{s}(t_0)^2 + \ddot{s}(t_0)^2 )\, ]^{\frac{1}{2}} + \O{\e^4},
  \nonumber \\
  \theta_{k,k+1} (t_0) & = & \frac{2\pi}{3} + \O{\e^2},\ \mbox{for $k
    \ne N-1$}  \nonumber \\
  \theta_{k,k+m} (t_0) & = & \frac{\pi}{2} + \O{\e^2},\ \mbox{for $m
    \ge 2$} \nonumber \\
  \theta_{k,N} (t_0) & = & \frac{\pi}{2} + \O{\e^2},
\end{eqnarray}
for $k = 1,\cdots,N-1$.

The angles between the neighbors are not orthogonal with each other.
That is based on the fact that $\vec{X}_k$ is not orthogonal to
$\vec{X}_{k+1}$, and is not essential.
Although one can let them orthogonal by changing the matrix
(\ref{eq:LinTrans}) into others, the notations will become messy.

The expression (\ref{eq:thetaNNgive}) and (\ref{eq:angle}) imply that
$N$ of the angles between the stable and unstable manifolds,
i.e. $\theta_{j,j},\, j = 1,\cdots,N$ are independent and sufficient
to write down the intersection of these $N$-dimensional manifolds, and
the results tell one that the only one angle is exponentially small
and the others are power of $\e$.
In the outer approximation (an approximation of the difference
equation (\ref{eq:DifferenceEq}) by the the differential equation with 
the singular perturbations (\ref{eq:ODE})), the outer equation
(\ref{eq:ODE}) forms a Hamiltonian system with $N$ degrees of
freedom in order-by-order and possesses an ``energy function.''
Therefore trajectories of the outer equation are restricted in
($2N-1$)-dimensional energy surface, and in the energy surface the
intersection of the $N$-dimensional stable and unstable manifolds
has $1$-dimension.
In reality, because we consider not continuous Hamiltonian systems but 
symplectic mappings, then such energy functions do not exist and the
separatrix splits into two different manifolds, i.e., the
$1$-dimensional stable and unstable manifold, in the exponentially
small order.\cite{HK}
We think that this feature is quite generic over the symplectic
mappings.
The naive approximation misses the exponentially small terms and leads 
to an incorrect result that the intersection angle between
sub-manifolds is equal to zero.
The approximation is however sufficient to estimate the angles of the
directions out of the separatrix.

Intersection angles are related with flux near stable and unstable
manifolds.
We expect that escape rate of trajectories near the manifolds gets
large as $N$ increases because the number of the angles in the power
of $\e$ dominates overwhelmingly.
We would like to confirm this conjecture with numerical experiments
in the future work.
This point is closely related with well-known Nekhoroshev's method and
a very important problem for understanding the global structure of
phase space of nearly integrable Hamiltonian systems with many degrees
of freedom.